\documentclass[twocolumn,showpacs,preprintnumbers,amsmath,amssymb]{revtex4}

\usepackage{graphicx}
\usepackage{dcolumn}
\usepackage{bm}

\begin{document}

\preprint{}

\title{Temporal and spectral disentanglement of \\laser-driven electron tunneling emission from a solid}

\author{Hirofumi Yanagisawa${}^{1,2}$}
\email{hirofumi@phys.ethz.ch}
\author{Sascha Schnepp${}^{3}$}
\author{Christian Hafner${}^{3}$}
\author{Matthias Hengsberger${}^{2}$}
\author{Alexandra Landsman${}^{1}$}
\author{Lukas Gallmann${}^{1,4}$}
\author{J\"{u}rg Osterwalder${}^{2}$}
\affiliation{${}^{1}$\mbox{Institute for Quantum Electronics, ETH Z\"{u}rich, CH-8093 Z\"{u}rich, Switzerland }\\
${}^{2}$\mbox{Physik-Institut, Universit\"{a}t Z\"{u}rich, CH-8057 Z\"{u}rich, Switzerland }\\
${}^{3}$\mbox{Laboratory for Electromagnetic Fields and Microwave Electronics, CH-8092 Z\"{u}rich, Switzerland} \\
${}^{4}$\mbox{Institute of Applied Physics, University of Bern, CH-3012 Bern, Switzerland}}

\begin{abstract}
By measuring energy spectra of the electron emission from a sharp tungsten tip induced by few-cycle laser pulses, the laser-field dependence of the emission mechanism was investigated. In strong laser fields, we confirm the appearance of laser-driven tunneling emission and find that it can be disentangled from the concomitant photo-excited electron emission, both temporally and spectrally, by the opening of a peculiar emission channel. This channel involves prompt laser-driven tunneling emission and subsequent laser-driven electron re-scattering off the surface, delayed by the electrons traveling far inside the metal before scattering. The quantitative understanding of these processes gives insights on attosecond tunneling emission from solids and should prove useful in designing new types of pulsed electron sources.
\end{abstract}

\date{\today}
\maketitle


A number of studies have clarified the intriguing characteristics of the electron emission processes when illuminating a nano-sized metallic tip with ultrashort laser pulses \cite{hommelhoff06b, hommelhoff10, hommelhoff11, hommelhoff12, ropers10,ropers12, lienau12,lienau13, Ang12, Ang13, Lee73, hommelhoff06a,  wu08, yanagisawa09, yanagisawa10, yanagisawa11, yanagisawa12,barwick07}. Plasmonic effects enhance optical electric fields at the tip apex \cite{hecht05}, showing spatial confinement and control of the electron emission on a nanometer scale \cite{hommelhoff06a, yanagisawa09, yanagisawa10}. The strength of the field at the emission site determines the emission mechanism, and the temporal confinement of the emitted electron pulses depends on the mechanism. For relatively weak fields, electrons excited by multi-photon absorption tunnel through the surface barrier or are emitted over the barrier as illustrated in Model A in Fig. 1 \cite{Ang12, Ang13, Lee73, hommelhoff06a,  wu08, yanagisawa09, yanagisawa10, yanagisawa11, yanagisawa12,barwick07}. These processes are insensitive to the laser phase and generate femtosecond electron pulses \cite{wu08, Ang12,yanagisawa11, yanagisawa12}. On the other hand, very strong fields largely modify the surface barrier and drive direct tunneling emission through the barrier as shown in Model B, producing attosecond coherent electron waves \cite{hommelhoff06b,Ang12}. Thus this laser-driven tunneling emission - also termed optical field emission - has become the subject of intense research in ultrafast science \cite{hommelhoff06b, hommelhoff10, hommelhoff11, hommelhoff12, ropers10,ropers12, lienau12,lienau13, Ang12, Ang13, Lee73, hommelhoff06a,  wu08, yanagisawa09, yanagisawa10, yanagisawa11, yanagisawa12,barwick07}. 

\begin{figure}[b]
\begin{center}
\includegraphics[scale=0.15]{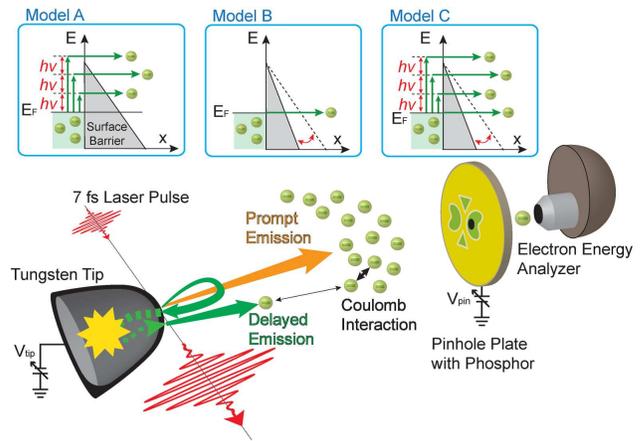} 
\end{center}
\vskip -\lastskip \vskip -3pt
\caption{\label{fig:epsart}
Conceptual diagrams of the laser-induced electron emission and experimental setup. A tungsten tip oriented towards the [011] crystal direction is mounted in a vacuum chamber. 7 fs laser pulses of 830 nm center wavelength are focused onto the apex of the tip. A pinhole plate covered with a phosphor is installed in front of the tip to observe the field emission pattern and to define the emission sites for spectroscopic measurements. A hemispherical electrostatic analyzer was used to measure the energy spectra. Different emission processes under strong fields are depicted by orange and green arrows. The insets show different emission mechanisms, referred to as Models A to C, discussed in this work.}
\label{fig:label-1}
\end{figure}

However, the photo-excited electron emission as in Model A is also enhanced in strong fields \cite{hommelhoff10,hommelhoff11,hommelhoff12}, and the two different mechanisms are difficult to distinguish experimentally due to laser-driven electron dynamics after emission  \cite{hommelhoff12}. Although some of the previous works explained their observations in terms of tunneling emission, the different mechanisms were never carefully disentangled \cite{ ropers12, lienau12, lienau13}. Therefore the emission mechanism in strong fields remains a hotly debated topic. Here, analyzing the laser-power dependence of emission spectra based on the three model scenarios in Fig. 1, we confirm the appearance of the laser-driven tunneling emission. It is characterized by the opening of a peculiar emission channel and can be disentangled from the other emission mechanisms temporally and spectrally. Our findings will enable the isolated study of laser-driven tunneling emission.

Measuring energy spectra of the emitted electrons is the most direct route to understanding the emission mechanism. We have therefore measured the energy spectra of the electron emission from a tungsten tip apex induced by ultrashort laser pulses, using an experimental setup as schematically drawn in Fig. 1. The energy spectra obtained in the strong-field regime show two characteristics: a plateau region spreading over several to tens of eV, similar to earlier observations \cite{hommelhoff11,hommelhoff12,ropers12, lienau12,lienau13}, and a prominent peak at low energy. Our quantitative simulations reveal that the plateau and peak features originate from different emission processes well separated in time, which we refer to as prompt and delayed emission, respectively. The prompt emission leads to a dense electron cloud as indicated by the orange arrow in Fig. 1. Strong space charge effects within this cloud are responsible for the formation of the plateau feature. Electrons emitted with a delay can avoid most of this strong Coulomb interaction in the vacuum and therefore pile up in a pronounced low-energy peak.

The delayed emission process is illustrated by green arrows in Fig. 1. Some of the emitted electrons are driven back to the surface by the oscillating laser field and re-scatter at the surface \cite{hommelhoff11,hommelhoff12, ropers12, lienau12, lienau13}. So far, the re-scattering process has been treated as instantaneous in the relevant literature \cite{hommelhoff11,hommelhoff12, ropers12,lienau12,lienau13}. However, the electrons travel typically a distance of the order of twice the mean free path inside the metal before reappearing in the vacuum after an elastic or inelastic back-scattering process (Fig. 1) \cite{pendry75}. At electron energies of the order of 5 to 10 eV, mean free path lengths are ranging from 7.5 nm to 12 nm \cite{lemell09}, which can lead to delays of the order of a few to even tens of femtoseconds with respect to the prompt emission. The number of electrons that are subject to this delayed emission depends strongly on the initial emission mechanism. Our simulations allow to distinguish between model scenarios A to C depicted in Fig. 1. The delayed emission channel becomes strong only for the laser-driven tunneling emission process, namely for Model B or C.

\begin{figure}[h!]
\begin{center}
\includegraphics[scale=0.3]{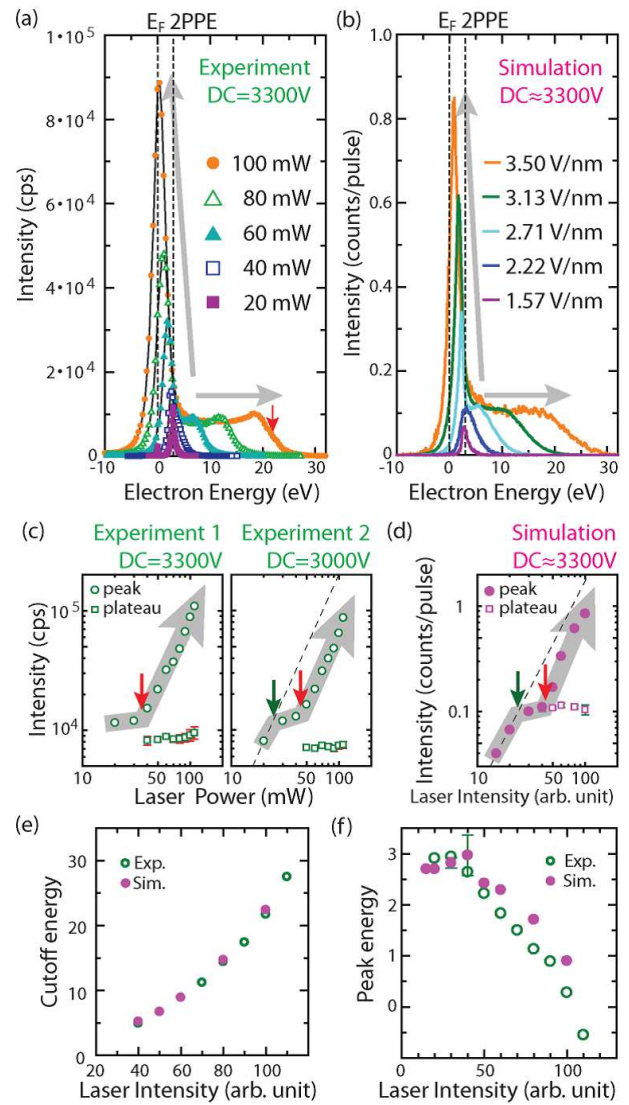} 
\end{center}
\vskip -\lastskip \vskip -3pt
\caption{\label{fig:epsart}
(a) Experimental and (b) simulated energy spectra for different laser intensities. Laser fields in the simulations are determined by first fitting the simulated spectrum to the experimental one at 60 mW with varying laser fields and then scaling the laser field properly with the laser powers in the experiments. They are defined by the maximum laser fields at the focus without field enhancement. The enhancement factor is 2.4 in our case. The Fermi energy $E_{F}$ is defined as 0 eV after subtraction of the DC bias voltage. (c) Evolution of the peak and plateau intensities in the experimental energy spectra for two different DC voltages and (d) the same for the simulated energy spectra. The dashed lines in (c) and (d) indicate lines with a slope of 2. (e) The cutoff energies of the plateaus and (f) the energies of the low-energy peak maxima from the experimental and simulated energy spectra are plotted as a function of laser power. An example for the cutoff energy of the plateau is indicated by a red arrow in (a). In (c) to (f), error bars are added for the case that the errors are bigger than symbols.}
\label{fig:label-2}
\end{figure}

The experimental energy spectra for different laser powers are shown in Fig. 2(a), where a smooth transition of the emission mechanism from the weak-field to the strong-field regime is observed. At the lowest laser power, a peak associated with electron emission from two-photon photoexcitation (2PPE) dominates the spectrum, which is typical for the weak-field regime \cite{yanagisawa11, yanagisawa12}. With increasing laser power, the peak grows in intensity and its maximum gradually moves towards lower energies. Concurrently a plateau feature with nearly constant intensity grows sideways towards higher energies. Gray arrows indicate these two general trends. This behaviour is totally different from that in the weak-field regime \cite{yanagisawa11, yanagisawa12}. This remarkable evolution of the peak feature has not been reported before, not even in strong-field experiments \cite{hommelhoff10,hommelhoff11,hommelhoff12,ropers12,lienau12, lienau13}. To analyze its evolution further, the peak and plateau intensities are plotted as a function of laser power in a log-log plot in the left panel of Fig. 2(c). At low laser powers where the 2PPE spectrum dominates, the initial slope is rather gentle but grows steeply after a pronounced upward kink indicated by a red arrow. In the right panel of the figure, another data set for a slightly lower DC voltage is shown. In this case, the tendency is similar, but the 2PPE peak shows two kinks (green and red arrows), with a steeper rise at the lowest laser powers. If the peak simply originated from the 2PPE excitations, the evolution would show a straight line with a slope of 2 as indicated by a dashed line. Therefore, the  kinks are strong indicators of changes in the emission process.

An intuitive interpretation of the spectra, however, is very difficult because of the space charge effects; approximately 1000 electrons per pulse were observed in measurements of the total emission current at the highest laser power. Hence, to understand the physics of the observed data, we have simulated electron trajectories in the vacuum with taking the space charge effects into account \cite{comment1}. All simulations were done in the full three dimensional system covering the geometry of the tip in front of the pinhole plate. An overview of the simulation methods is described below. 

There are four steps in the simulation. In the first step, solving Maxwell equations based on the Multiple Multipole Program \cite{openmax}, we simulated the time evolution of the optical local fields on the tip apex when the 7 fs laser pulse is passing by the apex of the tungsten tip. Secondly, the simulated local fields on the tip apex were used to calculate the emission current distribution on the apex as was done previously by using the Fowler-Nordheim theory \cite{yanagisawa09,yanagisawa10,yanagisawa11,yanagisawa12}. To calculate emission currents, the three emission models in Fig. 1 were used and compared. Thirdly, the electron trajectory simulation was performed. The total yield is obtained by integrating the calculated emission current over energy, time and space. The number of electrons per pulse was 850 at the highest laser intensity. The initial electron emission times, energies, positions and directions were determined based on the Monte Carlo method by using the calculated emission current distribution. The tunneling emission was delayed by the Keldysh time multiplied by a factor $C_{\tau}$ \cite{mcdonald13}. $C_{\tau}$ was used as a fitting parameter and adjusted to have the largest number of electrons re-directed to the tip; the resulting $C_{\tau}$ was 1.4. This maximization was found to yield the best agreement in the peak to plateau ratio of the simulated electron spectra as compared to the experiments. After determining the initial conditions, the electrons were propagated through the vacuum, where they feel four kinds of forces: 1. laser fields, 2. DC fields, 3. Coulomb forces between electrons (space charge effects), and 4. image charge forces. The electrons which are re-scattered from the tip experience the delay processes as illustrated in Fig. 1. Here we assume that 45 \% of the electrons can penetrate deeper into the surface and contribute to the delayed emission current while the rest of the electrons are removed from the computational field; this parameter was introduced to have intensities of the peak feature similar to the observations. The delay time was estimated by dividing the electron traveling length by the group velocity for the round trip from and back to the surface. To determine the path length which electrons travel before the inelastic scattering takes place, we assume that the scattering events are stochastic. The probability distribution follows the Poisson distribution with its maximum positioned at the inelastic mean free path (IMFP) of tungsten \cite{osterwalder90}. The IMFP is obtained from Ref. \cite{lemell09}. The thus calculated delay time is multiplied with a factor $C_{l}$ to have an evolution of the peak feature similar to the observations. The resulting $C_{l}$ was 1.8. This high value is considered to be due to the group velocity reaching only half of that of free electrons at around 5 eV above the Fermi level \cite{lemell09}. As a result, we obtained a mean total delay time of approximately 18 fs. The tracking-electron simulation stops when all the emitted electrons reach the counter electrode (pinhole plate). This process was repeated until enough statistics was obtained. In the forth step, the energy spectra are calculated from the final velocity of the electrons which enter the pinhole of the electrode like in our experimental setup in Fig. 1. Thus simulated spectra reveal the essential details of the experimental observations.

\begin{figure}[t]
\begin{center}
\includegraphics[scale=0.18]{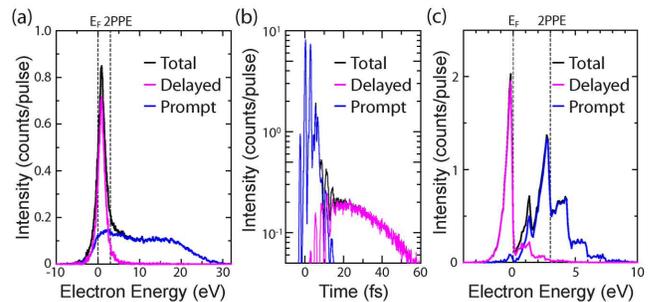}
\end{center}
\vskip -\lastskip \vskip -3pt
\caption{\label{fig:epsart}
(a) Simulated energy spectra at the counter electrode, (b) temporal profiles of the electron emission and (c) initial energy spectra on the tip apex decomposed into prompt and delayed emission processes. The laser field is 3.5 V/nm.}
\label{fig:label-3}
\end{figure}

The simulated energy spectra based on Model C are shown in Fig. 2(b). The plateau and the peak features are almost quantitatively reproduced. As explained above, these two features originate from the prompt and the delayed emission, respectively. Fig. 3(a) shows the simulated energy spectrum for the maximum laser field decomposed into contributions from the two emission processes. Clearly, the plateau represents the prompt emission and the peak the delayed emission. The two different features can be intuitively understood in terms of the relative strength of the space charge effects from a temporal profile of the electron emission as shown in Fig. 3(b). It shows an intense prompt emission happening within the first 10 fs, following closely the oscillations of the laser field, and a much weaker delayed emission during the next tens of femtoseconds. Both emission currents occur within a $30 \times 30$ nm$^2$ area \cite{yanagisawa09, yanagisawa10}. Therefore, the prompt emission is spatially and temporally dense enough to produce the broadened energy distribution due to the strong space charge effects \cite{supp2}. It is important to emphasize that avoiding the strong space charge effects of the prompt emission is essential to reproduce the peak feature; any electron emission process that occurs within the first 10 fs will only contribute to the plateau feature. Hence the low-energy peak in our measured data provides evidence for a delayed emission process. Our simulations based on Model C naturally include emission with the proper temporal delay and broadening, and therefore largely reduced space charge effects \cite{supp3}.

The simulations reveal an intuitive picture of the emission mechanisms. The initial energy distribution on the tip surface, before electron propagation through the vacuum, was decomposed into contributions from the prompt and the delayed emission processes as shown in Fig. 3(c). The delayed emission mainly consists of the laser-driven tunneling emission from the Fermi level, $E_{F}$. As will be discussed later, the number of electrons undergoing re-collision with the surface and their impact energies, forced by the oscillating laser field, should be highest for the laser-driven tunneling emission. In the delayed emission process, the electrons traveling in the metal after re-entry will typically lose 50 percent of their energy in an inelastic scattering event \cite{ritchie65,bauer02}. Therefore, the impact energy has to be more than approximately twice the barrier height for the inelastically back-scattered electrons to reappear in the vacuum. Because the impact energy depends on the exact time of the initial emission relative to the phase of the driving laser field \cite{shafir12}, only the laser-driven tunneling emission can produce a significant number of re-colliding electrons with high enough energies owing to its strong dependence on the laser oscillation phase  \cite{hommelhoff06b, Ang12}. Photo-excited electron emission like in Model A, on the other hand, is rather insensitive to the laser phase \cite{Ang12}, leading to only few electrons with high impact energy. Therefore,  Model A can be excluded as it cannot produce the low-energy peak. Note that also Model B, considering exclusively laser-driven tunneling emission, can reproduce the spectra in the strong-field regime quite well, but Model C leads to generally more quantitative agreement over the whole range of the transition regime; the 2PPE peak at the lowest laser power will never be reproduced by Model B.

The simulations clarify also the physics of the observed kinks in the peak evolution in Fig. 2(c). Fig. 2(d) shows the same plot resulting from the simulated spectra. There are two kinks in the evolution similar to the experimental data. The first kink at the green arrow turns out to be an intensity saturation of the 2PPE peak due to the space charge effects \cite{wendelen12}. After saturation, the emission from the 2PPE and higher multi-photon excitation starts to expand sideward as the plateau feature. On the other hand, after the second kink at the red arrow, the peak intensity increases again. This signals the opening of the delayed emission channel due to the laser-driven tunneling emission, as the impact energies of the electrons on the surface become higher than twice the barrier height \cite{supp5}. In addition, the energy location of the plateau cutoff and the peak position in the spectra are also successfully reproduced by our simulations as shown in Figs. 2(e) and (f). The excellent agreement of the plateau cutoff energies indicates the successful modeling of the space charge effects and electron photo-excitation. The shift of the low-energy peak originates from the remaining weaker Coulomb interaction between the prompt and the delayed electrons; the slight disagreement is due to the simplified description in our model of the complex physics. 

In summary, we have observed a delayed emission channel opening in strong fields that is clearly associated with laser-driven electron tunneling, mainly from the Fermi level. The delayed emission manifests itself by a prominent low-energy peak. Thus the laser-driven tunneling emission is disentangled from the photo-excited electron emission temporally and spectrally. This delayed emission provides information on the attosecond dynamics of electron tunneling at a solid surface, as well as on the femtosecond scale electron scattering processes within the tip material. Moreover, the quantitative understanding of these processes as demonstrated here should be useful in designing electron sources with attosecond temporal confinement. One of the more daring ideas is to use a carbon nanotube with a closed end on one side and an open end on the other side. Illuminating the closed end with strong-field laser pulses should lead to the electron emission from the open end on the other side. The emission from the open end should be still in phase with the oscillation of the laser fields driving the electron emission because the electrons re-entering the nanotube would be transported ballistically. Thus temporal control of the electron emission with true attosecond resolution should be possible. Such a pulsed electron source should find applications in ultrafast science, particularly impacting cutting-edge technology that relies on bright and coherent electron beams like electron diffraction, microscopy, or holography \cite{Zuo03, wang00,tonomura05,gomer93}.\\

This work was supported by the Swiss National Science Foundation through the {\it Ambizione} (grant number PZ00P2\_131701) and the {\it NCCR MUST}, and Kazato Research Foundation. We appreciate that Dr. Christoph Lemell gave us his unpublished data of the simulated IMFP at low electron energies. We thank Prof. U. Keller, Prof. B. Rethfeld, Prof. M. Aeschlimann, Prof. C. Oshima, Prof. K. Watanabe, Prof. T. Greber, Dr. S. Tsuchiya, Dr. C. Cirelli and Dr. L. Castiglioni for fruitful discussions and M. Baer for technical support.


\end{document}